\documentclass[prl,twocolumn,showpacs,preprintnumbers,amsmath,amssymb]{revtex4}

\usepackage{graphicx}
\usepackage{dcolumn}
\usepackage{bm}

\begin{document}

\preprint{APS/123-QED}

\title{Enhanced superconducting gaps in the tri-layer high-$T_{\rm{c}}$ cuprate Bi$_2$Sr$_2$Ca$_2$Cu$_3$O$_{10+\delta}$}

\author{S. Ideta$^1$, K. Takashima$^1$, M. Hashimoto$^1$, T. Yoshida$^1$, A. Fujimori$^1$, H. Anzai$^2$, T. Fujita$^2$, Y. Nakashima$^2$, A. Ino$^2$, M. Arita$^3$, H. Namatame$^3$, M. Taniguchi$^{2, 3}$, K. Ono$^4$, M. Kubota$^4$, D. H. Lu$^5$, Z.-X. Shen$^5$, K. M. Kojima$^1$, S. Uchida$^1$}
\affiliation{$^1$Department of Physics, University of Tokyo, Bunkyo-ku, Tokyo 113-0033, Japan\\$^2$Graduate School of Science, Hiroshima University, Higashi-Hiroshima 739-8526, Japan\\$^3$Hiroshima Synchrotron Center, Hiroshima University, Higashi-Hiroshima 739-0046, Japan\\$^4$Photon Factory, Institute of Materials Structure Science, KEK, Tsukuba, Ibaraki 305-0801, Japan\\$^5$Department of Applied Physics and Stanford Synchrotron Radiation Laboratory, Stanford University, Stanford, CA94305
}

\date{\today}

\begin{abstract}
We have investigated the optimally doped tri-layer cuprate Bi$_2$Sr$_2$Ca$_2$Cu$_3$O$_{10+\delta}$ (Bi2223) by angle-resolved photoemission spectroscopy, and  observed energy bands and Fermi surfaces originated from the outer and inner CuO$_2$ planes (OP and IP) separately. The OP band is overdoped (hole density $\sim$ 0.26/Cu) with a large $d$-wave gap of $\Delta_0$ $\sim$ 43 meV while the IP is underdoped ($\sim$ 0.06/Cu) with an even large gap of $\Delta_0$ $\sim$ 60 meV. We propose that the enhancement of the $\Delta_0$ of IP is due to the minimal influence of out-of-plane disorder and that the OP gap is then enhanced through interlayer coupling, most likely a proximity effect from the IP.
\end{abstract}

\pacs{Valid PACS appear here}
\maketitle

It has been well known that one of the most efficient ways to increase the critical temperature ($T_{\rm{c}}$) of high-$T_{\rm{c}}$ cuprate superconductors (HTSCs) is to increase the number of neighboring CuO$_2$ planes ($n$). $T_{\rm{c}}$ generally increases from single-layer ($n$ = 1), double-layer ($n$ = 2), to tri-layer ($n$ = 3) and then decreases for $n$ $\geq$ 4 \cite{Iyo0}. So far, several mechanisms have been proposed to explain the $n$ dependence of $T_{\rm{c}}$. According to the tunneling mechanism of Cooper pairs between the CuO$_2$ planes, $T_{\rm{c,max}}$ should increase with increasing $n$ \cite{Wheatley}. However if one takes into account the charge imbalance between the planes and the existence of competing order, $T_{\rm{c,max}}$ takes a maximum at $n$ = 3 \cite{S.Chakravarty} in agreement with experiment. Meanwhile, $T_{\rm{c}}$ shows tendency to increase with next-nearest-neighbor Cu-Cu hopping parameter $t^{\prime}$, which increases with the number of CuO$_2$ planes \cite{Pavarini}. Also, $T_{\rm{c}}$ increases with decreasing degree of out-of-plane disorder \cite{Eisaki, Fujita}. So far, which governs the $n$ dependence of $T_{\rm{c,max}}$ has been unclear because of the lack of detailed knowledge about the electronic structure of the multi-layer cuprates.

In the case of Bi-based HTSCs, the optimum $T_{\rm{c}}$ ($T_{\rm{c,max}}$) increases from the single-layer Bi$_2$Sr$_2$CuO$_{6+\delta}$ (Bi2201, $T_{\rm{c,max}}$ = 35 K), the double-layer Bi$_2$Sr$_2$CaCu$_2$O$_{8+\delta}$ (Bi2212, $T_{\rm{c,max}}$ = 95 K) to the tri-layer Bi$_2$Sr$_2$Ca$_2$Cu$_3$O$_{10+\delta}$ (Bi2223, $T_{\rm{c,max}}$ = 110 K). Angle-resolved photoemission spectroscopy (ARPES) studies of double-layer Bi2212 \cite{Feng2, Yamasaki} and four-layer Ba$_2$Ca$_3$Cu$_4$O$_8$F$_2$ (F0234) \cite{Chen} have revealed the splitting of band dispersions and Fermi surfaces (FSs). In Bi2212, hybridization between the two CuO$_2$ planes causes splitting into the bonding and anti-bonding bands \cite{Feng2}. The ARPES study on F0234 has indicated band splitting due to the different hole concentrations of the outer CuO$_2$ planes and the inner CuO$_2$ planes, and correspondingly two FS sheets have been observed \cite{Chen}. The maximum superconducting (SC) gap was $\sim$ 60 meV, approximately twice as large as that of the smaller one. On the other hand, previous ARPES results on the tri-layer Bi2223 have not revealed band splitting and multiple FSs \cite{Sato2, Feng1, Matsui}. In the present work, we have successfully observed the band splitting of Bi2223 through a detailed photon-energy dependent study and indicated that the outer CuO$_2$ plane (OP) and the inner CuO$_2$ plane (IP) have different doping levels and gap magnitudes. We can thus correlate the high $T_{\rm{c,max}}$ in Bi2223 with the large SC gaps for the OP and IP.
 
Single crystals of optimally doped Bi2223 ($T_{\rm{c}}$ = 110 K) were grown by the travelling solvent floating zone (TSFZ) method \cite{Fujii}. ARPES experiments were carried out at BL 9A of Hiroshima Synchrotron Radiation Center ($h\nu$ = 6.6 - 12.9 eV, circularly polarized light), BL 5-4 of Stanford Synchrotron Radiation Laboratory ($h\nu$ = 18.5 eV, linear polarized light), BL 28A of Photon Factory ($h\nu$ = 45 eV), and the University of Tokyo (He I$\alpha$ $h\nu$ = 21.218eV). The total energy resolution ($\Delta E$) was set at 5, 11, 15 and 18 meV, respectively. The samples were cleaved $in\ situ$ under an ultrahigh vacuum of $\sim$ 1$\times$10$^{-11}$ Torr. The temperature of experiments were performed at $T$ = 10 K.
\begin{figure}[ht]
\includegraphics[width=7.9cm]{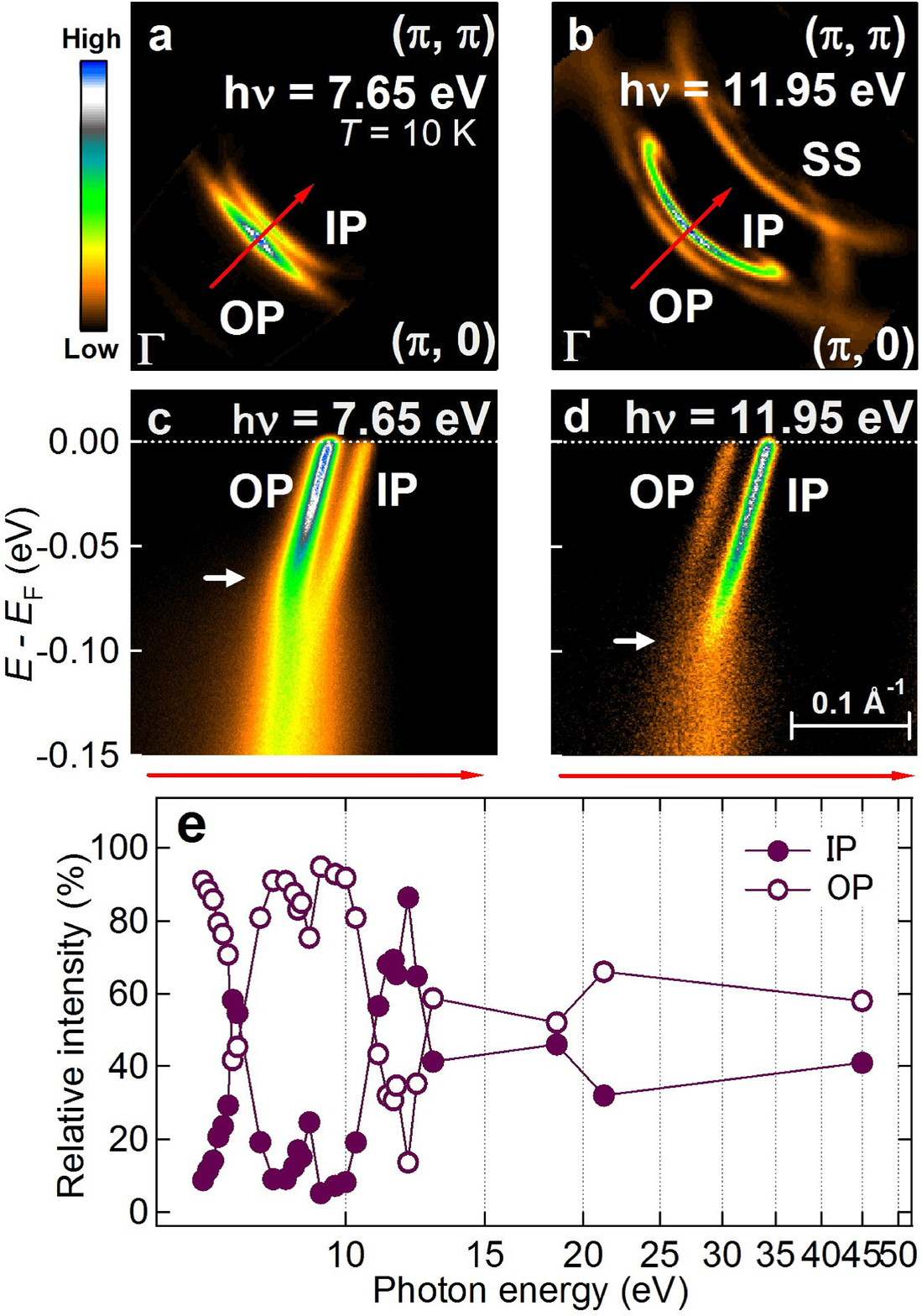}
\caption{(Color online) (a), (b): Intensity plots of ARPES spectra for Bi2223 at $E$ = $- 40 \pm 5$ meV in momentum space. Two Fermi surfaces are observed corresponding to the outer CuO$_2$ plane (OP) and inner CuO$_2$ plane (IP). Superstructures due to the Bi-O layer modulation are indicated by SS. (c), (d): Band dispersions in the nodal direction. Red arrows in (c) and (d) correspond to the nodal direction as shown by red arrows in (a) and (b), respectively. The energies of the kinks for the OP and IP bands are different as shown by white arrows. (e): Relative intensities of the OP and IP bands as functions of photon energy. For $h\nu$ = 7.65 eV [(a), (c)], the OP band spectra are enhanced while for $h\nu$ = 11.95 eV [(b), (d)], the IP band is enhanced. Measurements have been performed at $T$ = 10 K.}
\end{figure}

In Figs. 1(a) and 1(b), we show spectral weight mapping at 40 meV below the Fermi level ($E_{\rm{F}}$) integrated within a window $\pm$ 5 meV taken with $h\nu$ = 7.65 and 11.95 eV, respectively. Two FS sheets corresponding to the OP and IP are successfully resolved, and the intensity of OP is enhanced for $h\nu$ = 7.65 eV while that of IP is enhanced for $h\nu$ = 11.95 eV. As shown in Fig. 1(e), the intensity ratio of the OP and IP bands deduced from the momentum distribution curve's (MDC) at $E_{\rm{F}}$ in the nodal direction are indeed strongly dependent on photon energy. In particular, in the low-photon energy region, it changes dramatically with small changes in photon energy. Here, we assign the FS sheet closer to the $\Gamma$ point to the OP, and the other FS sheet to the IP, following the NMR studies of tri-layer HTSCs [(Cu,C)Ba$_2$Ca$_3$Cu$_3$O$_{10-\delta}$ (Cu1223), HgBa$_2$Ca$_2$Cu$_3$O$_{8+\delta}$ (Hg1223), etc], where the hole concentration of the OP is found to be larger than that of the IP \cite{Kotegawa2}. 

The dispersions in the nodal direction corresponding to Figs. 1(a) and 1(b) are shown in Figs. 1(c) and 1(d), respectively. Since there are three neighboring CuO$_2$ planes, one would expect to observe three bands which come from two OP and one IP bands, as indicated by $t$-$t^{\prime}$-$t^{\prime\prime}$-$J$ model calculation \cite{Mori}. The fact that we  observed only two bands implies that the two OP bands, i.e., the bonding and anti-bonding OP bands, are nearly degenerate. In fact, the full width at half maximum (FWHM) of the MDC for the OP band at $E_{\rm{F}}$, $\sim$ 0.011 ($\rm{\AA^{-1}}$), is significantly larger than that of the IP band $\sim$ 0.0074 ($\rm{\AA^{-1}}$), suggesting that two unresolved OP bands exist. The FWHM of the IP band is nearly the same as that of the bonding or anti-bonding band in Bi2212, 0.0065 ($\rm{\AA^{-1}}$) \cite{Yamasaki}, indicating its single-component nature. The FSs for the OP and IP have been determined by the minimum-gap-locus in the SC state. The values of the tight-binding (TB) parameter are $-t'/t$ $\sim$ 0.26 (OP) and $\sim$ 0.29 (IP), if the third-nearest-neighbor hopping parameter is assumed to be $-t^{\prime\prime}/t^{\prime}$ = 0.5. The hole concentration for the OP and IP bands deduced from the FS areas are $\sim$ 26 and $\sim$ 6 $\%$, respectively. These values are quantitatively different from those estimated from the NMR results \cite{Kotegawa2} for the optimally doped Hg1223, i.e., $\sim$ 0.35 (OP) and $\sim$ 0.30 (IP). The discrepancy is probably due to hybridization between the OP and IP, which reduces the \textquotedblleft net\textquotedblright\ charge differences probed by NMR compared with the \textquotedblleft nominal\textquotedblright\ charge differences probed by ARPES.

In Figs. 2(a) and 2(b), we show dispersions in the SC state for the OP and IP bands taken with $h\nu$ = 9 and 11.95 eV, respectively, from the nodal to off-nodal cuts. The gap energies for both bands \begin{figure}[ht]
\includegraphics[width=9.3cm]{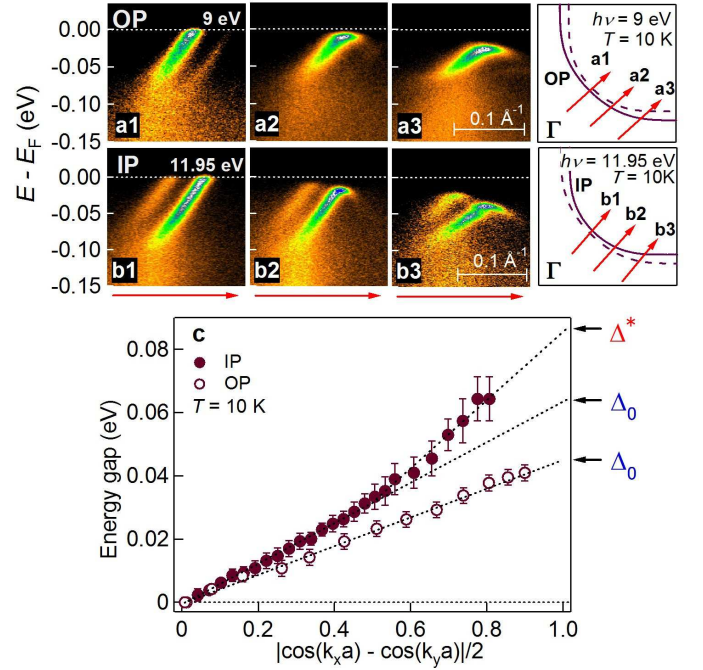}
\caption{(Color online) Energy-momentum intensity plots for cuts from the nodal to the off-nodal regions for the OP band [(a1) - (a3)] and for the IP band [(b1) - (b3)] observed in the superconducting (SC) state ($T$ = 10 K). The corresponding cuts are shown in the right panels. The photon energies, $h\nu$ = 9 eV and 11.95 eV, enhance the OP and IP bands, respectively. (c): Momentum dependences of the OP and IP bands. The definition of the SC gap $\Delta_{0}$ and the anti-nodal gap $\Delta^{*}$ is shown.}
\end{figure}\begin{figure}[ht]\includegraphics[width=8cm]{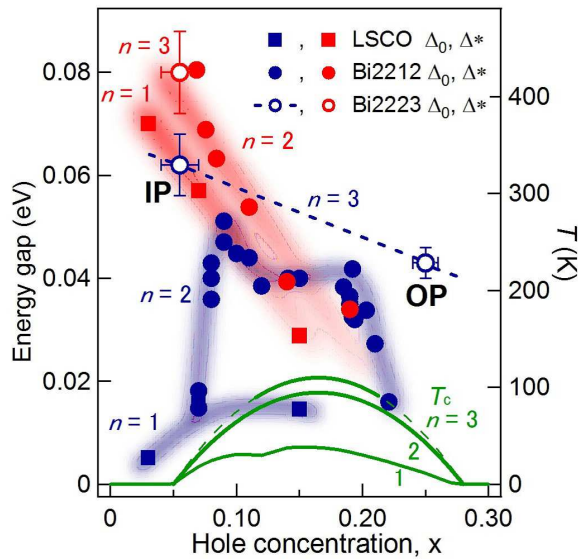}
\caption{(Color online) Comparison of $\Delta_0$ and $\Delta^*$ between the single-, double-, and tri-layer cuprates as functions of hole concentration. Their $T_{\rm{c}}$ are plotted by solid curves. One can see that the $\Delta_0$'s of LSCO, Bi2212 and Bi2223 show a strong dependence on the number of CuO$_2$ plane $n$ while $\Delta^*$ only weak one. The $\Delta_0$ and $\Delta^*$ values for $n$ = 1 and 2 have been taken from ARPES results for LSCO \cite{Yoshida2} and Bi2212 \cite{Tanaka, Lee, Ding1, Mesot, Feng2}. Red and blue shades are guide to the eye.}
\end{figure} are very different, as in the case of F0234 ($n$ = 4). The gap magnitudes for the OP and IP bands have been estimated from the peak position of the symmetrized energy distribution curve (EDC) at each $k_{\rm{F}}$ fitted to the phenomenological model \cite{Norman}. The momentum dependence of the gap magnitude for OP is almost simple $d$-wave, $\Delta_0|\cos(k_xa) - \cos (k_ya)|/2$ with $\Delta_0$ $\sim$ 43 meV, as shown by a straight line in Fig. 2(c). On the other hand, the gap for the IP band deviates from the simple $d$-wave around the anti-node $\sim(\pi$, 0). The gap size is characterized by two parameters $\Delta_0$ $\sim$ 60 meV around the node and $\Delta^*$ $\sim$ 80 meV in the aniti-nodal region, where $\Delta_0$ and $\Delta^*$ are defined by the linear extrapolation of the gap magnitude to the anti-node ($|\cos (k_xa)- \cos (k_ya)|/2 = 1$), as shown in Fig. 2(c). Because the deviation of the gap anisotropy from the simple $d$-wave is known to be prominent in underdoped cuprates \cite{Tanaka, Lee}, the observed gap anisotropy is consistent with the doping levels of the OP and IP estimated from the FS areas. Judging from the present result on the gap of OP, $\Delta_0$ $\sim$ 43 meV, one can conclude that the previous ARPES result $\Delta_0$ $\sim$ 40 meV for Bi2223 reflected the OP band due to the employed photon energies of $\sim$ 22 eV \cite{Feng1, Matsui, Sato1, Sato2}. 

In Fig. 3, $\Delta_0$ and $\Delta^*$ for La$_{2-x}$Sr$_x$CuO$_4$ (LSCO, $n$ = 1), Bi2212 ($n$ = 2) and Bi2223 ($n$ = 3) are plotted as functions of hole doping. In spite of its heavily overdoping, the $\Delta_0$ of the OP band is almost the same as that of optimally doped Bi2212. Furthermore, the $\Delta_0$ of the IP is  much larger than that of underdoped Bi2212.  The interpolated values of the $\Delta_0$ of the OP and IP to the optimal doping ($x$ $\sim$ 0.16), $\sim$ 50 meV, is larger than $\Delta_0$ $\sim$ 40 meV of Bi2212 in the optimally doped region by $\sim$ 20 $\%$, which ratio almost coincides with the $T_{\rm{c}}$ ratio between Bi2223 and Bi2212. One can see that the $\Delta_0$ for the same doping level increases with increasing $n$ from $n$ = 1 to $n$ = 3 while $\Delta^*$ weakly increases. Thus, we conclude that the larger size of $\Delta_0$ leads to the higher $T_{\rm{c}}$ of $\sim$ 110 K in Bi2223 than those in LSCO and Bi2212.

Now, let us discuss possible origins of the enhancement of $\Delta_0$ in Bi2223 compared with those in the single- and double-layer cuprates shown in Fig. 3. We may consider three possible origins that could lead to the large $\Delta_0$ for both OP and IP: (1) The values of the next-nearest-neighbor hopping parameter $- t^{\prime}/t$ which has correlation with $T_{\rm{c,max}}$  \cite{Pavarini}, $\sim$ 0.26 for OP and $\sim$ 0.29 for IP, is larger than those of the single-layer ($- t^{\prime}/t$ $\sim$ 0.15 - 0.2) \cite{Yoshida3, Hashimoto2} and double-layer ($- t^{\prime}/t$ $\sim$ 0.24) \cite{Lee3} cuprates; (2) Interlayer tunneling of Cooper pairs between OP and IP may enhance the SC order parameter \cite{S.Chakravarty}; (3) Out-of-plane disorder effect is small in IP because IP is protected from the out-of-plane disorder by the OP, and therefore, IP is ideally flat \cite{Eisaki}. In Fig. 4, we test the correlation between $T_{\rm{c,max}}$, SC gap $\Delta_0$, $- t^{\prime}/t$ at optimal doping ($x$ $\sim$ 0.16) \footnote{For $n$ = 3, values interpolated between IP and OP have been used.}, and $\Delta^{*}$ in the underdoped region ($x$ $\sim$ 0.06) \footnote{We have plotted the $\Delta^*$ values for LSCO and Bi2212 on the red shades at $x$ $\sim$ 0.06 in Fig. 3.} for the single-layer (Bi2201, LSCO), double-layer (Bi2212) and tri-layer (Bi2223) HTSCs. One can see a remarkable correlation between $\Delta_0$ and $T_{\rm{c,max}}$, that is, $T_{\rm{c,max}}$ is nearly proportional \begin{figure}[ht]
\includegraphics[width=7.5cm]{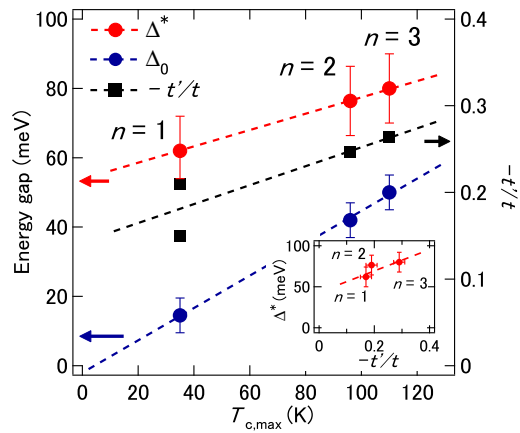}
\caption{(Color online) Correlation between $\Delta_0$, $-t^{\prime}/t$ at optimal doping ($x$ $\sim$ 0.16), $\Delta^*$ in the underdoped region ($x$ $\sim$ 0.06), and $T_{\rm{c,max}}$ from the single- ($n$ = 1) to tri-layer ($n$ = 3) cuprates. $\Delta_0$ shows the strongest correlation with $T_{\rm{c,max}}$ ($T_{\rm{c,max}}$ $\propto$ $\Delta_0$) while $-t^{\prime}/t$ and $\Delta^*$ show only weak correlation with $T_{\rm{c,max}}$.  Parameter values of $-t^{\prime}/t$ for $n$ = 1 and 2 have been taken from ARPES results for LSCO \cite{Yoshida3}, Bi2201 \cite{Hashimoto2} and Bi2212 \cite{Lee3}. Inset shows correlation between $\Delta^*$ and $-t^{\prime}/t$ at $x$ $\sim$ 0.06. The $-t^{\prime}/t$ values for $n$ = 1, 2 in the inset are taken from Ref.\cite{Yoshida3} for LSCO and Refs.\cite{Lee3, Kordyuk} for Bi2212.}
\end{figure}to $\Delta_0$. On the other hand, $T_{\rm{c,max}}$ and $\Delta^*$ shows only weak correlation compared with $T_{\rm{c,max}}$ $versus$ $\Delta_0$. $-t^{\prime}/t$ shows some correlation with $T_{\rm{c,max}}$, but not so strong as $T_{\rm{c,max}}$ $versus$ $\Delta_0$.\\
\ \ \ In this context, it is interesting to point out that $\Delta^*$ is weakly dependent on $-t^{\prime}/t$, and therefore the large $\Delta^*$ of IP can be explained by its large $-t^{\prime}/t$ value as shown in the inset of Fig. 4 \footnote{As for the $-t^{\prime}/t$ of Bi2212 in the inset of Fig. 4, we have plotted the extrapolated value at $x$ $\sim$ 0.06 from $-t^{\prime}/t$ $\sim$ 0.24 at $x$ $\sim$ 0.16 (Ref.\cite{Lee3}) and $\sim$ 0.20 at $x$ $\sim$ 0.1 (Ref.\cite{Kordyuk}).}. Such a dependence of $\Delta^*$ on $-t^{\prime}/t$ has been predicted by a $t$-$t^{\prime}$-$J$ model calculation \cite{Prelovsek}. The out-of-plane disorder may enhance $\Delta^*$ \cite{Hashimoto3, Okada}, however, the large $\Delta^*$ of IP cannot be explained by this effect since IP is protected from the out-of-plane disorder. \\
\ \ \ In Bi2223, $\Delta_0$ for IP is very large possibly due to the protection from the out-of-plane disorder by the presence of OPs. $\Delta_0$ for OP is also larger than that in Bi2212 despite the influence of the out-of-plane disorder to the same extent as in the case of Bi2212. This unusually large $\Delta_0$ of OP can be explained by a proximity effect from IP having the very large $\Delta_0$. Such a proximity effect would in turn reduce the $\Delta_0$ of IP. The fact that the $\Delta_0$ of IP is nevertheless very large means that the \textquotedblleft original\textquotedblright $\Delta_0$ of IP was even larger, or that the proximity effect did not reduce the $\Delta_0$ of IP through a mechanism proposed by Okamoto and Maier \cite{Okamoto}: According to their calculation, the combination of overdoped and underdoped CuO$_2$ planes can enhance the SC order parameter of the overdoped planes while the SC order parameter of the underdoped plane remains unchanged. It may also be possible that the interlayer tunneling of Cooper pairs within the tri-layer \cite{S.Chakravarty} enhances the SC order parameter of the tri-layer system, although interlayer Cooper pair tunneling observed by optical conductivity measurement is known to have negligible effect on $T_{\rm{c}}$ \footnote{J.~Sch$\rm{\ddot{u}}$tzmann, H.~S.~Somal, A.~A.~Tsvetkov, D.~van~der~Marel, G.~E.~J. Koops, N.~Koleshnikov, Z.~F.~Ren, J.~H.~Wang, E.~Br$\rm{\ddot{u}}$ck, and A.~A.~Menovsky, Phys. Rev. B \textbf{55}, 11118 (1997).}. Further experimental and theoretical studies are necessary to clarify these possibilities. \\
\ \ \ In conclusion, we have measured ARPES spectra of the optimally doped Bi2223 and investigated the precise electronic structure of the outer and inner CuO$_2$ planes and Fermi surfaces. We find that the OP and IP have different hole concentrations, and that $\Delta_0$ for the IP band is very large probably due to the negligible influence of out-of-plane disorder. The relatively large $\Delta_0$ of the OP in spite of its overdoping may originate from the proximity effect from the underdoped IP planes. \\
\ \ \ We thank C. Panagopoulos, T. P. Devereaux, S. Okamoto, and Y. Mori for informative discussion. ARPES experiments were carried out at HiSOR, Hiroshima Synchrotron Radiation Center, Hiroshima University (Proposal No. 07-A-10), at KEK-PF (Proposal No. 2006S2-001) and at SSRL, which is operated by a Grant-in-Aid for Scientific Research in Priority Area gInvention of Anomalous Quantum Materialsh.


\end{document}